\documentclass[nofootinbib,twocolumn,pra,showpacs,english,aps]{revtex4}
\usepackage[T1]{fontenc}
\usepackage[latin9]{inputenc}
\usepackage{color}
\usepackage{array}
\usepackage{amsmath}
\usepackage{graphicx}
\usepackage{amssymb}
\usepackage{epstopdf}
\usepackage{caption}
\usepackage{epsfig}
\usepackage{bm}
\usepackage{epsfig}
\usepackage{color}
\usepackage{subfig}
\usepackage{babel}

\makeatletter

\@ifundefined{definecolor}
 {
}{}
\def\be{\begin{equation}}
\def\ee{\end{equation}}
\def\bea{\begin{eqnarray}}
\def\eea{\end{eqnarray}}

\@ifundefined{showcaptionsetup}{}{ \PassOptionsToPackage{caption=false}{subfig}}
\makeatother

\begin{document}

\title{{\normalsize Higher-order nonclassical properties of atom-molecule
Bose-Einstein Condensate}}

\author{{\normalsize Sandip Kumar Giri$^{1,2}$, Kishore Thapliyal$^{3}$,
Biswajit Sen$^{4}$, Anirban Pathak$^{3,}$\footnote{Email: anirban.pathak@gmail.com,  Phone: +91 9717066494}%
}}

\affiliation{$^{1}$Department of Physics, Panskura Banamali College, Panskura-721152,
India\\
$^{2}$Department of Physics, Vidyasagar University, Midnapore-721102,
India \\
$^{3}$Jaypee Institute of Information Technology, A-10, Sector-62,
Noida, UP-201307, India\\
$^{4}$Department of Physics, Vidyasagar Teachers' Training College,
Midnapore-721101, India\\
}
\begin{abstract}
{\normalsize The transient quantum statistical properties of the atoms
and molecules in an atom-molecule BEC system are investigated by obtaining
a third-order perturbative solution of the Heisenberg's equations
of motion corresponding to the Hamiltonian of an atom-molecule BEC
system where two atoms can collide to form a molecule. Time dependent
quantities like two boson correlation, entanglement, squeezing, antibunching,
etc., are computed and their properties are compared. It is established
that atom-molecule BEC system is highly nonclassical as lower-order
and higher-order squeezing and antibunching in pure (atomic and molecular)
modes, squeezing and antibunching in compound mode and lower-order
and higher-order entanglement in compound mode can be observed in
the atom-molecule BEC system. Exact numerical results are also reported
and analytic results obtained using the perturbative technique are
shown to coincide with the exact numerical results.}{\normalsize \par}
\end{abstract}


\maketitle

\section{{\normalsize Introduction}}

Nonclassical properties of radiation field have been investigated
since long \cite{dodonov}. However, the interest on nonclassical
states has not been decreased over time. Interestingly, interest on
nonclassical states has been considerably increased in recent past
as several applications of nonclassical states are recently reported
in the context of quantum computation and communication \cite{Ekert protocol,CV-qkd-hillery,Bennet1993,densecoding}.
Specifically, nonclassical states are shown to be essential for the
implementation of all the recently proposed protocols of device independent
quantum cryptography and a bunch of traditional protocols of discrete
\cite{Ekert protocol} and continuous variable quantum cryptography
\cite{CV-qkd-hillery}, quantum teleportation \cite{Bennet1993},
dense-coding \cite{densecoding}, etc. Further, the focus of study
of nonclassical properties has recently been extended beyond quantum
optics (i.e., beyond nonclassical properties of radiation field),
and nonclassical properties are recently been investigated in several
atomic {[}\cite{pathak-pra2} and references therein{]} and optomechanical
systems \cite{optomechanical}. To a large extent, these studies are
motivated by the atom-optics analogy, and the fact that recently several
possibilities of implementation of quantum computing devices using
superconductivity-based systems \cite{super-cond-system-1,scalable quantum circuit},
and Bose Einstein condensate (BEC) based systems \cite{josephson-qubit,macroscopic qc,quantum information transfer,metrology1,metrology2}\textcolor{red}{{}
}have been reported. For example, two weakly coupled BECs confined
in a double-well trap is shown to produce Josephson charged qubits
\cite{josephson-qubit}; possibility of implementation of quantum
algorithms using BECs is reported \cite{macroscopic qc}; Josephson
qubits that are suitable for implementation of a scalable integrated
quantum circuits are realized \cite{scalable quantum circuit}; quantum
state is transferred using cavities containing two-component BECs
coupled by optical fiber \cite{quantum information transfer}; schemes
for implementing protocols of quantum metrology using two-component
BECs are proposed \cite{metrology1,metrology2}. Thus, many of the
recently reported applications of BECs in quantum information processing
involve two-mode BECs (\cite{quantum information transfer,metrology1,metrology2}
and references therein). These facts have motivated us to systematically
investigate the nonclassical properties of a specific class of two-mode
BEC systems that are usually referred to as atom-molecule BEC system
\cite{Vardi-2001}. Here it would be apt to note that the two-mode
BECs can be broadly classified into two classes: (i) atom-atom BEC
where total number of bosons present in the system is conserved \cite{josephson-qubit,macroscopic qc,quantum information transfer,Bec-experiment,he2,Jing,ion-trap,hine1,he1,he3,fan},
and (ii) atom-molecule BEC where total number of bosons present in
the system is not conserved as two or more bosons of atomic mode can
combine to form a boson in molecular mode and equivalently a boson
in molecular mode can decompose into two or more bosons in atomic
mode \cite{Vardi-2001,hine1,IJTP-2009-dai,Jin}.

A state is called nonclassical if its Glauber-Sudarshan $P$ function
is not a classical probability density, which implies that for a nonclassical
state the $P$ function is either negative or more singular than $\delta$
function. This $P$ function based definition of nonclassicality is
both necessary and sufficient, but $P$ function is not directly measurable
through experiments. Because of this experimental restriction, several
other operational criteria of nonclassicality are proposed, which
are sufficient but not necessary. For example, zeroes of $Q$ function,
negativity of Wigner function, Fano factor, $Q$ parameter, various
inseparability criteria, etc., are introduced as operational criteria
for detection of nonclassicality. Here we restrict ourselves to a
set of nonclassical criteria that are experimentally realizable and
can characterize a set of nonclassical characters of practical importance,
such as, antibunching, intermodal antibunching, squeezing, intermodal
squeezing, intermodal entanglement, etc. Further, recently higher-order
nonclassical properties are experimentally observed in some bosonic
systems \cite{Maria-PRA-1,Maria-2,higher-order-PRL}. In these experimental
works, it has been clearly observed that it may be easier to detect
a weak nonclassicality when we use a criterion of higher-order nonclassicality
(cf. Fig. 4 of \cite{Maria-PRA-1}). Earlier,\textcolor{red}{{} }the
existence of higher-order nonclassicality was theoretically studied
in a number of quantum optical systems \cite{HOAwithMartin,HOAis not rare,generalized-higher order}.
However, except two very recent works \cite{pathak-pra2,BEC with perinova},
not much effort has been made until now to obtain the signatures of
higher-order nonclassicalities in the coupled BEC systems. Keeping
these facts in mind, we aim to study the possibilities of observing
higher-order squeezing, higher-order antibunching and higher-order
entanglement in two-mode atom-molecule BEC system. Present study on
nonclassicality in atom-molecule BEC is also motivated by the fact
that the nonclassical characters investigated here are already shown
to be useful for various important tasks related to quantum communication. 

Although the basic concept of BEC was known since 1924 \cite{bose-bec,einstein},
it has been experimentally observed only in mid-nineties \cite{Bec-experiment}.
Since then the interest on BEC has been considerably increased, and
it has been observed in various systems. For example, BEC is observed
in ultra-cold dilute alkali gases using magneto-optical traps (\cite{ion-trap}
and references therein). These demonstrations and amplified interest
on BEC lead to a set of recent studies on nonclassical properties
of two-mode BECs {[}\cite{pathak-pra2,Vardi-2001,he2,hine1,he1,he3,fan,BEC with perinova}
and references therein{]}\textcolor{blue}{.}\textcolor{red}{{} }A two-mode
BEC system may be visualized as a combined system of two BECs where
each mode is a BEC and a particular boson can only occupy one of the
two modes. However, bosons from one mode can shift to the other mode.
As described above, mainly two types of two-mode BECs exist, namely
atom-atom BEC and atom-molecule BEC. Recently, we have systematically
studied nonclassical properties of atom-atom BEC \cite{pathak-pra2},
and the present work aims to extend that to two-mode atom-molecule
BEC system. There exist several Hamiltonians for the two-mode atom-molecule
BEC system, most of them are equivalent. Here we restrict ourselves
to a specific Hamiltonian that was introduced by Vardi \cite{Vardi-2001},
and was recently used by Perinova \textit{et al.} \cite{BEC with perinova}
to investigate the existence of nonclassical states. Equations of
motion corresponding to this two-mode BEC Hamiltonian can be solved
by using different approaches, such as the short-time approximation
\cite{short-time}, Gross-Pitaevski mean-field theory \cite{GPE},\textcolor{red}{{}
}Perinova et al.'s invariant subspace method \cite{BEC with perinova},
etc. In the present work, a third order analytic operator solution
of the two-mode atom-molecule BEC Hamiltonian of our interest is obtained
by using a perturbative technique developed by some of the present
authors \cite{bsen1,bsen2,bsen3,bsen4,bsen5}. We have also obtained
exact numerical solution. The third order perturbative solutions obtained
here as the\textcolor{red}{{} }time evolution of the annihilation and
creation operators of different bosonic modes are subsequently used
to illustrate the existence of various types of lower-order, and higher-order
nonclassicality in the two-mode atom-molecule BEC system. Specifically,
existence of lower-order, and higher-order squeezing and antibunching
in pure (atomic and molecular) modes, squeezing in compound mode and
lower-order, and higher-order entanglement in compound mode are shown
in the atom-molecule BEC system.

Remaining part of the present paper is organized as follows. In Sec.
\ref{sec:Model-Hamiltonian}, we briefly introduce the model Hamiltonian
that describes the two-mode atom-molecule BEC system. We also report
a third-order perturbative solution of the Heisenberg's equations
of motion corresponding to the field modes present in this Hamiltonian.
In Sec. \ref{sec:Quadrature-squeezing}, we illustrate the existence
of squeezing of quadrature variables for the individual and coupled
modes of the two-mode atom-molecule BEC system. Existence of higher-order
squeezing using Hillery's criterion of amplitude powered squeezing
is also shown. Similarly, in Sec. \ref{sec:Photon-antibunching},
it is shown that the antibunching can be observed for all the individual
and coupled modes of the two-mode BEC system studied in this paper.
Existence of higher-order antibunching is also shown in the individual
modes. In Sec. \ref{sec:Entanglement}, lower-order, and higher-order
quantum entanglement is studied using a set of inseparability criteria
and intermodal entanglement is observed.\textcolor{red}{{} }Finally,
we conclude the paper in Sec. \ref{sec:Conclusions}.

\section{{\normalsize Model Hamiltonian\label{sec:Model-Hamiltonian}}}

The Hamiltonian of the atom-molecule BEC is given by \cite{Vardi-2001}

\begin{equation}
\begin{array}{lcl}
H & = & \frac{\hbar\Delta}{2}a^{\dagger}a+\frac{\hbar\Omega}{2}\left(a^{\dagger}a^{\dagger}b+aab^{\dagger}\right),\end{array}\label{eq:hamiltonian}\end{equation}
where $a\,(a^{\dagger})$ and $b\,(b^{\dagger})$ are the annihilation
(creation) operators for the atomic and molecular modes, respectively.
Bosons present in each mode constitute a BEC. Further, in this model
two bosons in atomic mode (i.e., two atoms) can combine to form a
boson in molecular mode (i.e., a molecule) without the generation
of heat, and the atomic mode is coupled to the molecular mode by the
near resonant two boson transition or Feshbech resonance where the
detuning is $\Delta$, and the coupling constant is $\Omega$. In
order to study the lower-order, and higher-order nonclassicalities
in this atom-molecule BEC system, we first construct the Heisenberg's
equations of motion for atomic and molecular mode as \begin{equation}
\begin{array}{lcl}
\dot{a}(t) & = & -i\left(\frac{\Delta}{2}a(t)+\Omega a^{\dagger}(t)b(t)\right),\\
\dot{b}(t) & = & -\frac{i\Omega}{2}a^{2}(t).\end{array}\label{eq:eq of motion}\end{equation}
This set of coupled nonlinear differential equations are not exactly
solvable in closed analytical form. Here we use a perturbative approach
developed by Sen and Mandal \cite{bsen1,bsen2,bsen3,bsen4,bsen5}.
It is already established (see \cite{pathak-pra2} and references
therein) that the perturbative solutions obtained by Sen-Mandal approach
are more general than the solutions obtained by conventional short-time
approximation\textcolor{red}{{} }\cite{short-time}. In this approach
the solution of Eq. (\ref{eq:eq of motion}) is assumed as follows 

\begin{equation}
\begin{array}{lcl}
a(t) & = & f_{1}a(0)+f_{2}a^{\dagger}(0)b(0)+f_{3}a^{\dagger}(0)a^{2}(0)\\
 & + & f_{4}a(0)b^{\dagger}(0)b(0)+f_{5}a^{\dagger}(0)b(0)+f_{6}a^{\dagger}(0)b^{\dagger}(0)b^{2}(0)\\
 & + & f_{7}a^{\dagger2}(0)a(0)b(0)+f_{8}a^{3}(0)b^{\dagger}(0),\\
b(t) & = & g_{1}b(0)+g_{2}a^{2}(0)+g_{3}b(0)+g_{4}a^{\dagger}(0)a(0)b(0)\\
 & + & g_{5}a^{2}(0)+g_{6}a^{\dagger2}(0)b^{2}(0)+g_{7}a^{2}(0)b^{\dagger}(0)b(0)\\
 & + & g_{8}a^{\dagger}(0)a^{3}(0).\end{array}\label{eq:evolution}\end{equation}
The parameters $f_{i}$ and $g_{i}$ (where $i\in\left\{ 1,2,\cdots,8\right\} $)
are evaluated with the help of the boundary condition $f_{1}(0)=g_{1}(0)=1$
and $f_{i}(0)=g_{i}(0)=0$\textcolor{green}{{} }for $i\in\left\{ 2,3,\ldots,8\right\} $.
Under these initial conditions, we obtain solutions for $f_{i}(t),$
and $g_{i}(t)$ as \begin{equation}
\begin{array}{lcl}
f_{1} & = & e^{-\frac{i\triangle t}{2}},\\
f_{2} & = & -\frac{2i\Omega}{\triangle}\sin\frac{\triangle t}{2},\\
f_{3} & = & -\frac{f_{4}}{2}=\frac{i\Omega^{2}}{\triangle^{2}}\left(\sin\frac{\triangle t}{2}-\frac{\triangle t}{2}f_{1}\right),\\
f_{5} & = & -\frac{f_{6}}{2}=\frac{f_{7}}{3}=-\frac{i\Omega^{3}}{\triangle^{3}}\left(\triangle t\cos\frac{\triangle t}{2}-2\sin\frac{\triangle t}{2}\right),\\
f_{8} & = & -\frac{i\Omega^{3}}{\triangle^{3}}f_{1}\left(\triangle t-\sin\triangle t\right),\\
g_{1} & = & 1,\\
g_{2} & = & \frac{f_{1}f_{2}}{2},\\
g_{3} & = & \frac{g_{4}}{2}=f_{1}f_{3}^{*},\\
g_{5} & = & -\frac{g_{7}}{4}=\frac{g_{8}}{2}=\frac{f_{1}f_{5}}{2},\\
g_{6} & {\normalcolor =} & f_{1}f_{8}^{*}.\end{array}\label{eq:solution}\end{equation}
The above solution is valid up to the third order in $\Omega$\textcolor{green}{{}
}with $\Omega t<1$, such that the perturbation theory is respected.
In what follows, we have used these solutions to study the possibilities
of observing various nonclassical effects in the two-mode atom-molecule
BEC system described by (\ref{eq:hamiltonian}). Further, to investigate
the signatures of nonclassical characters of the two-mode atom-molecule
BEC system, we consider that initially the atomic mode and the molecular
mode are coherent. Thus, at $t=0$, the composite state of the system
can be viewed as a product (separable) state $|\alpha\beta\rangle=|\alpha\rangle\otimes|\beta\rangle,$
which is the product of two coherent states $\left|\alpha\right\rangle $
and $\left|\beta\right\rangle $ that are eigenkets of $a$ and $b$,
respectively. Therefore, we can write \begin{equation}
\begin{array}{lcl}
a(0)\left|\alpha,\beta\right\rangle  & = & \alpha\left|\alpha,\beta\right\rangle ,\,\,\, b(0)\left|\alpha,\beta\right\rangle =\beta\left|\alpha,\beta\right\rangle .\end{array}\label{eq:coherent state}\end{equation}
In the following sections, we have investigated the existence of various
types of lower-order, and higher-order nonclassicalities in the atom-molecule
BEC system described by (\ref{eq:hamiltonian}) using the solution
reported in (\ref{eq:evolution})-(\ref{eq:solution}) and the initial
state $|\alpha\beta\rangle$.

\section{{\normalsize Quadrature squeezing\label{sec:Quadrature-squeezing}}}

\begin{widetext}

\begin{figure}[h]
\begin{centering}
\includegraphics[scale=0.55,angle=-90]{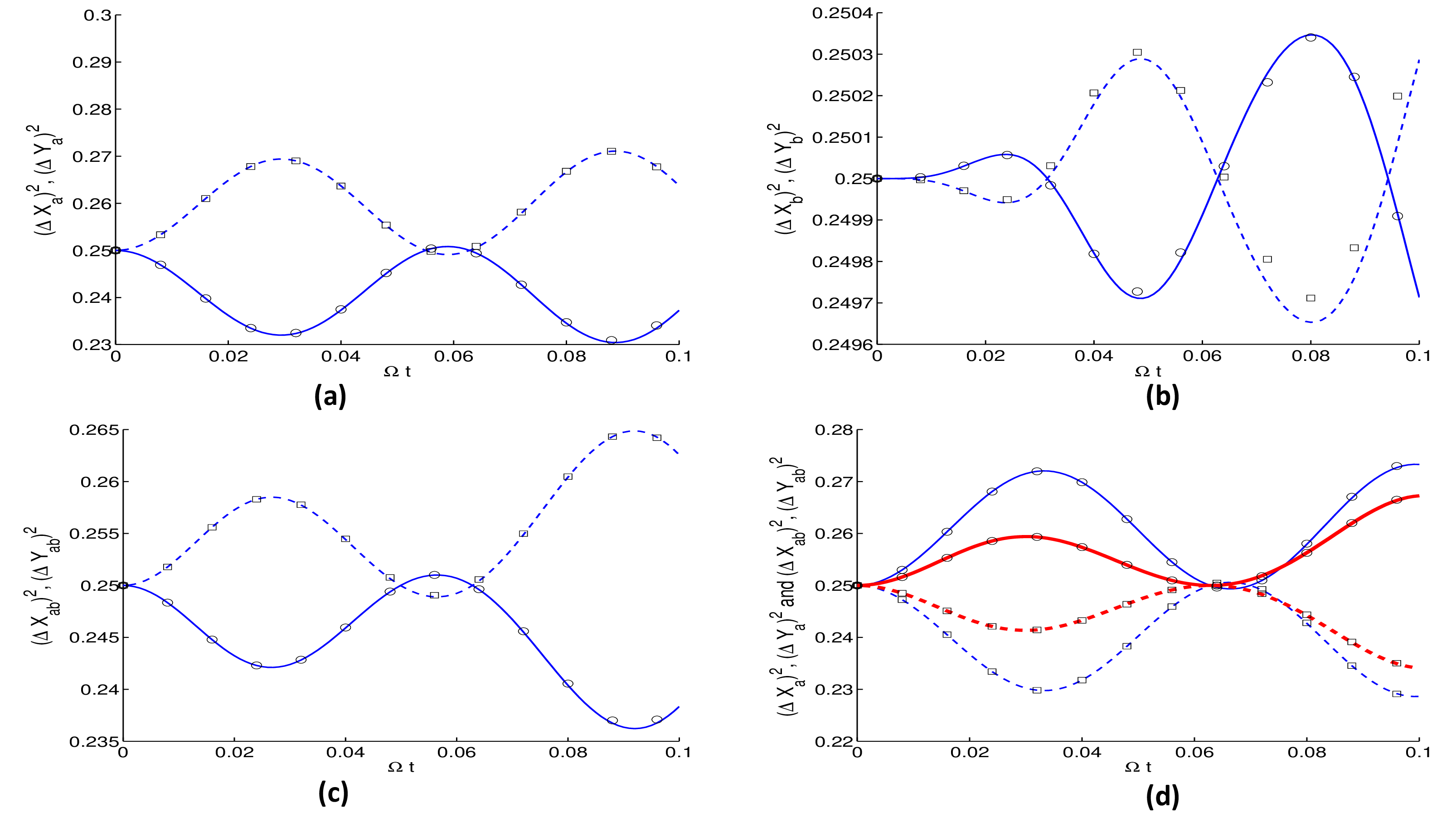}
\par\end{centering}

\caption{\label{fig:squeezing}(Color online) Quadrature fluctuation with rescaled
interaction time $\Omega t$ of (a) mode $a$ (b) mode $b$, and (c)
coupled mode $ab$ for $\Omega=10^{2},\,\frac{\Delta}{\Omega}=10^{2},\,\alpha=5,$
and $\beta=2$. In (a) and (b) smooth (dashed) line denotes variance
in $X_{i}\,\left(Y_{i}\right)$ quadrature where $i\in\{a,b\}$, and
in (c) smooth (dashed) line represents variance in $X_{ab}\,\left(Y_{ab}\right)$
quadrature using approximate analytic solution. Circle (square) represents
the variance using exact numerical solution for respective modes.
In (d), quadrature fluctuation of atomic mode $a$ (thin blue lines)
and compound atom-molecule mode $ab$ (thick red lines) are shown
for $\Omega=10^{2},\,\frac{\Delta}{\Omega}=10^{2},\,\alpha=5,$ and
$\beta=-2$. (d) shows that the phase of the input coherent state
can be used to control the amount of squeezing.}
\end{figure}

\end{widetext}

In order to investigate the quadrature squeezing effect in the atomic
and molecular modes in atom-molecule BEC, we use the standard definition
of quadrature operators. For example, in atomic mode quadrature operator
is defined as \begin{equation}
\begin{array}{lcl}
X_{a} & = & \frac{1}{2}\left(a(t)+a^{\dagger}(t)\right),\\
Y_{a} & = & \frac{1}{2i}\left(a(t)-a^{\dagger}(t)\right).\end{array}\label{eq:quadrature-a}\end{equation}
Similarly, we can construct the quadrature operators $X_{b}$ and
$Y_{b}$ for the molecular mode $b$. Further, the squeezing in the
compound mode $ab$, can be examined using quadrature operators for
the compound atomic-molecular mode as \begin{equation}
\begin{array}{lcl}
X_{ab} & = & \frac{1}{2\sqrt{2}}\left(a(t)+a^{\dagger}(t)+b(t)+b^{\dagger}(t)\right),\\
Y_{ab} & = & -\frac{i}{2\sqrt{2}}\left(a(t)-a^{\dagger}(t)+b(t)-b^{\dagger}(t)\right).\end{array}\label{eq:quadrature-ab}\end{equation}
Quadrature squeezing in $i$-th mode (where $i\in\left\{ a,b\right\} $)
is witnessed if the fluctuation in one of the quadrature operators
goes below the minimum uncertainty level, i.e., if we observe \begin{equation}
\left(\Delta X_{i}\right)^{2}<\frac{1}{4}\,\mathrm{\, or}\,\,\left(\Delta Y_{i}\right)^{2}<\frac{1}{4}.\label{eq:condition for squeezing}\end{equation}
Using (\ref{eq:evolution}), (\ref{eq:solution}), (\ref{eq:coherent state})
and (\ref{eq:quadrature-a}) we can obtain analytic expressions for
the fluctuation of the quadrature operators in atomic mode as \begin{equation}
\begin{array}{lcl}
\left[\begin{array}{c}
\left(\triangle X_{a}\right)^{2}\\
\left(\triangle Y_{a}\right)^{2}\end{array}\right] & = & \frac{1}{4}\left[1+2\left|f_{2}\right|^{2}\left|\beta\right|^{2}+\left(2f_{2}^{*}f_{3}\alpha^{2}\beta^{*}+{\rm c.c.}\right)\right.\\
 & \pm & \left\{ f_{1}f_{3}\alpha^{2}+\left(f_{1}f_{2}+f_{1}f_{5}\right)\beta+6f_{1}f_{5}\left|\alpha\right|^{2}\beta\right.\\
 & + & \left.\left.\left(f_{1}f_{6}+f_{2}f_{4}\right)\left|\beta\right|^{2}\beta+{\rm c.c.}\right\} \right],\end{array}\label{eq:squ-a}\end{equation}
where ${\rm c.c.}$ stands for the complex conjugate. Similarly, for
the pure mode $b$ and the coupled mode $ab$ the analytic expressions
of quadrature fluctuations can be obtained as follows \begin{equation}
\left[\begin{array}{c}
\left(\triangle X_{b}\right)^{2}\\
\left(\triangle Y_{b}\right)^{2}\end{array}\right]=\frac{1}{4}\left[1\pm\left\{ \left(g_{7}+2g_{2}g_{4}\right)\alpha^{2}\beta+{\rm c.c.}\right\} \right],\label{eq:squ-b}\end{equation}
and\begin{widetext} \begin{equation}
\begin{array}{lcl}
\left[\begin{array}{c}
\left(\Delta X_{ab}\right)^{2}\\
\left(\Delta Y_{ab}\right)^{2}\end{array}\right] & = & \frac{1}{4}\left[1+\left|f_{2}\right|^{2}\left|\beta\right|^{2}+\left(f_{2}^{*}f_{3}\alpha^{2}\beta^{*}+f_{2}g_{4}^{*}\left|\beta\right|^{2}\alpha{}^{*}+{\rm c.c.}\right)\pm\frac{1}{2}\left\{ f_{1}f_{3}\alpha^{2}\right.\right.\\
 & + & \left(f_{1}f_{2}+f_{1}f_{5}\right)\beta+2f_{1}g_{4}\alpha\beta+4f_{1}g_{5}\alpha^{3}+\left(f_{1}f_{6}+f_{2}f_{4}\right)\left|\beta\right|^{2}\beta\\
 & + & \left.\left.\left(g_{7}+2g_{2}g_{4}\right)\alpha^{2}\beta-4f_{8}\alpha^{*}\beta^{2}+6f_{1}f_{5}\left|\alpha\right|^{2}\beta+{\rm c.c.}\right\} \right].\end{array}\label{eq:squ-ab}\end{equation}
\end{widetext}Eqs. (\ref{eq:squ-a})-(\ref{eq:squ-ab}) indicate
that within the domain of validity of the approximate solution reported
here, quadrature squeezing may exist in atomic, molecular and compound
modes. To illustrate this, plots of right hand sides of Eqs. (\ref{eq:squ-a}),
(\ref{eq:squ-b}), and (\ref{eq:squ-ab}) are shown in Fig. \ref{fig:squeezing}
a, \ref{fig:squeezing} b and \ref{fig:squeezing} c, respectively.
Regions of these plots with variance less than $\frac{1}{4}$ explicitly
illustrate the existence of single mode quadrature squeezing in atomic
and molecular modes, and also intermodal squeezing in atom-molecule
coupled mode. Interestingly, when we consider $\alpha$ and $\beta$
as real and $\alpha=5$ and $\beta=2$, then quadrature squeezing
is dominantly observed in only one quadrature $X_{a}\,(X_{ab})$ in
atomic (compound) mode, and it is weakly observed in a small region
in the other quadrature, i.e., $Y_{a}\,(Y_{ab})$ (see Fig. \ref{fig:squeezing}
a (c)). However, it is clearly observed in both the quadratures in
molecular mode (see Fig. \ref{fig:squeezing} b). Interestingly, by
increasing the value of $\alpha$, amount of quadrature squeezing
in $Y_{a}$ and $Y_{ab}$, and corresponding region of nonclassicality
can be considerably increased. For example, for $\alpha=10$ and $\beta=2$
we can observe squeezing in $X_{a},\, Y_{a},\, X_{ab}\,{\rm and\,}Y_{ab}$
(not shown in the figure). Even for lower values of $\alpha$, large
squeezing in $Y_{a}\,(Y_{ab})$ can be obtained by using appropriate
choice of phase of initial coherent states. This fact is illustrated
in the Fig. \ref{fig:squeezing} d, where we have considered $\alpha=5$
and $\beta=-2$ (i.e., $\beta=2\exp(i\pi)$). In Fig. \ref{fig:squeezing}
d, squeezing in atomic mode (thin blue lines) and intermodal squeezing
in atom-molecule coupled mode (thick red lines) are plotted together
to show squeezing in $Y_{a}$ and $Y_{ab}$ with change of phase of
molecular mode. Further, the boxes and circles shown in the plots
depict exact numerical results obtained by integrating the time-dependent
Schrodinger equation corresponding to the Hamiltonian (\ref{eq:hamiltonian})
by defining the operators as matrices. The excellent coincidence of
the exact numerical results with the analytic results obtained using
the perturbative solution obtained here strongly establishes the accuracy
and validity of our perturbative solution. Interestingly, similar
coincidence of perturbative results with the exact numerical results
is also observed in other signatures of nonclassicalities that are
discussed in the present work. However, we have not shown the numerical
results in all cases.

\subsection{Higher-order squeezing}

\begin{figure}
\begin{centering}
\includegraphics[scale=0.35]{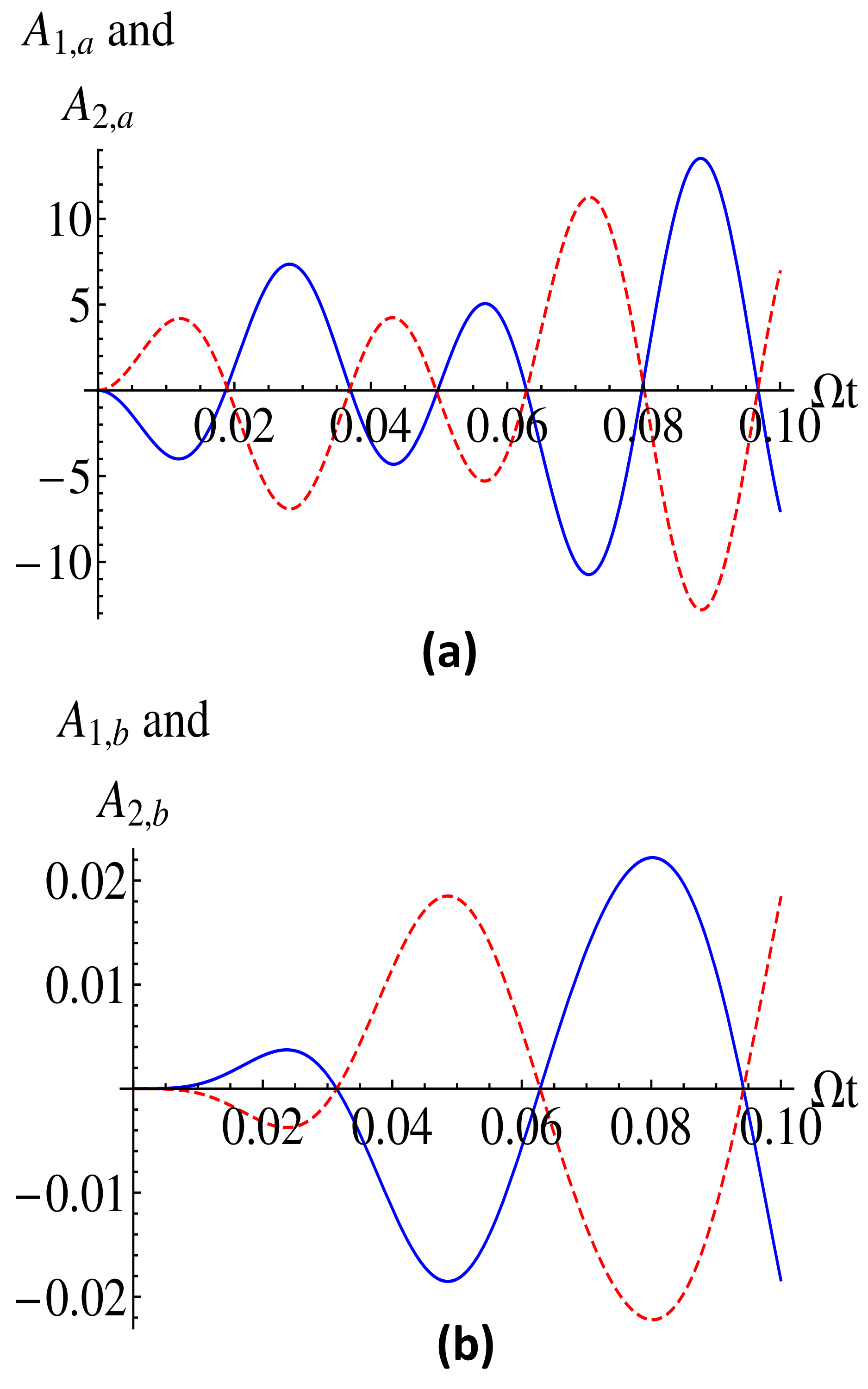}
\par\end{centering}

\caption{\label{fig:Amp-sq-squeezing}(Colour online) Amplitude squared squeezing
is observed in (a) atomic mode $a$, and (b) molecular mode $b$ for
$\Omega=10^{2},\,\frac{\Delta}{\Omega}=10^{2},\,\alpha=10,$ and $\beta=2$.
Negative parts of the smooth blue line show squeezing in quadrature
variable $Y_{1,a}\left(Y_{1,b}\right)$ and that of dashed red line
illustrate squeezing in quadrature variable $Y_{2,a}\left(Y_{2,b}\right)$.}
\end{figure}

So far, we have examined the existence of lower-order squeezing in
the two-mode atom-molecule BEC using (\ref{eq:evolution}), (\ref{eq:solution})
and (\ref{eq:condition for squeezing}). The same may be easily extended
to the higher-order squeezing that is usually studied using two alternative
approaches \cite{HIlery-amp-sq,Hong-Mandel1,HOng-mandel2}. Specifically,
higher-order squeezing is studied either using Hillery's criterion
of amplitude powered squeezing \cite{HIlery-amp-sq}, that provides
witness for the existence of higher-order nonclassicality through
the reduction of variance of an amplitude powered quadrature variable
for a quantum state with respect to its coherent state counterpart,
or using the criterion of Hong and Mandel \cite{Hong-Mandel1,HOng-mandel2},
which reflects the existence of higher-order nonclassicality through
the reduction of higher-order moments of the usual quadrature operators
with respect to their coherent state counterparts. In the present
paper we have restricted ourselves to Hillery's criterion. Specifically,
Hillery introduced amplitude powered quadrature variables as \begin{equation}
Y_{1,a}=\frac{a^{k}+\left(a^{\dagger}\right)^{k}}{2}\label{eq:quadrature-power1}\end{equation}
and \begin{equation}
Y_{2,a}=i\left(\frac{\left(a^{\dagger}\right)^{k}-a^{k}}{2}\right).\label{eq:quadrature-power2}\end{equation}
As $Y_{1,a}$ and $Y_{2,a}$ do not commute, we can easily obtain
a condition of squeezing. For example, for $k=2$, Hillery's criterion
for amplitude squared squeezing is \begin{equation}
A_{i,a}=\left\langle \left(\Delta Y_{i,a}\right)^{2}\right\rangle -\left\langle N_{a}+\frac{1}{2}\right\rangle <0,\label{eq:criterion-amplitude squared}\end{equation}
where $i\in\{1,2\}.$ Now using (\ref{eq:evolution}), (\ref{eq:solution}),
(\ref{eq:quadrature-power1}) and (\ref{eq:quadrature-power2}) we
obtain\begin{widetext}\begin{eqnarray}
\left(\begin{array}{c}
A_{1,a}\\
A_{2,a}\end{array}\right) & = & 2\left|f_{2}\right|^{2}\left|\alpha\right|^{2}\left|\beta\right|^{2}+\left\{ 2\left(f_{1}f_{5}^{*}+f_{1}^{*}f_{8}\right)\left|\alpha\right|^{2}\alpha^{2}\beta^{*}\right.\nonumber \\
 &  & \left.+2f_{1}f_{2}^{*}\left|f_{2}\right|^{2}\alpha^{2}\beta^{*}\left|\beta\right|^{2}-f_{2}^{*}f_{3}\alpha^{2}\beta^{*}+c.c.\right\} \pm\left\{ f_{1}^{3}f_{3}\alpha^{4}\right.\label{eq:Amp squ-a}\\
 &  & +f_{1}f_{2}^{3}\alpha^{*2}\beta^{3}+\frac{1}{2}f_{1}^{2}f_{2}^{2}\left(1+4\left|\alpha\right|^{2}\right)\beta^{2}+f_{1}^{2}\left(f_{1}f_{2}+7f_{1}f_{5}+f_{2}f_{3}\right)\alpha^{2}\beta\nonumber \\
 &  & \left.+2f_{1}^{2}\left(2f_{2}f_{3}+3f_{1}f_{5}\right)\left|\alpha\right|^{2}\alpha^{2}\beta+f_{1}^{2}\left(3f_{2}f_{4}-2f_{1}f_{5}\right)\alpha^{2}\left|\beta\right|^{2}\beta+c.c\right\} \nonumber \end{eqnarray}
 and\begin{equation}
\left(\begin{array}{c}
A_{1,b}\\
A_{2,b}\end{array}\right)=\pm\left\{ \left(g_{7}+2g_{2}g_{4}\right)\alpha^{2}\beta^{3}+{\rm c.c.}\right\} .\label{eq:Amp squ-b}\end{equation}
\end{widetext}Variation of right hand sides of (\ref{eq:Amp squ-a})
and (\ref{eq:Amp squ-b}) are plotted in Fig. \ref{fig:Amp-sq-squeezing}.
Existence of amplitude squared squeezing in both the quadratures of
atomic (molecular) mode can be clearly observed in Fig. \ref{fig:Amp-sq-squeezing}
a (b) where negative parts of the plots depict signature of higher-order
squeezing. In the present study, we have restricted ourselves to the
study of amplitude squared squeezing. However, it is possible to investigate
the existence of Hong-Mandel type higher-order squeezing, and amplitude
$k$-th power squeezing using the time evolution of the field operators
obtained here and the approach adopted above to study lower-order,
and higher-order squeezing.

\section{{\normalsize Antibunching\label{sec:Photon-antibunching}}}

\begin{widetext}

\begin{figure}[h]
\begin{centering}
\includegraphics[scale=0.5,angle=-90]{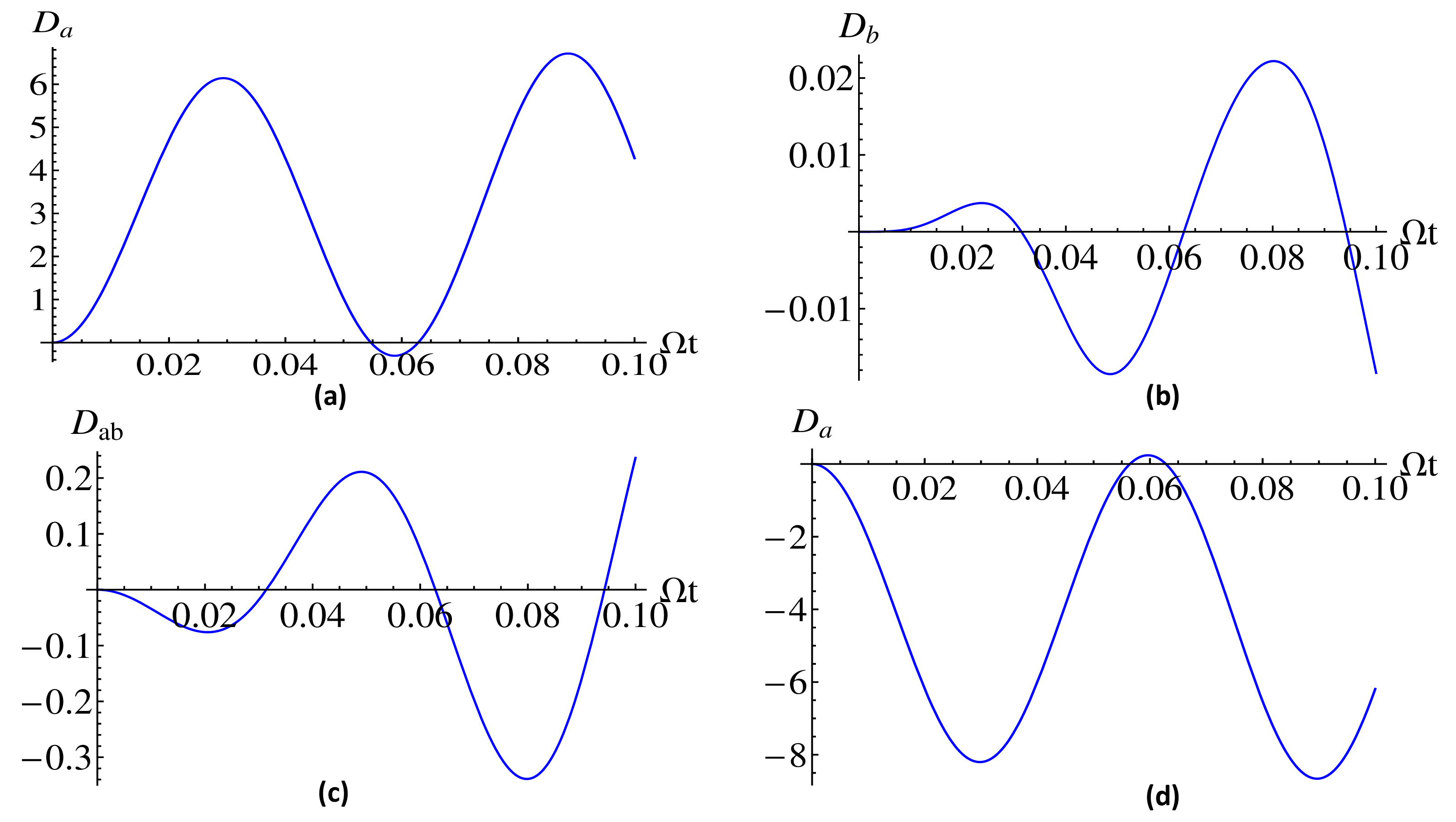}
\par\end{centering}

\caption{\label{fig:statistics}(Color online) Plot of $D_{i}$ with rescaled
interaction time $\Omega t$ for (a) mode $a,$ (b) mode $b$, and
(c) coupled mode $ab$ for $\Omega=10^{2},\,\frac{\Delta}{\Omega}=10^{2},\,\alpha=10,$
and $\beta=2$, and\textcolor{green}{{} }(d) mode $a$ for $\Omega=10^{2},\,\frac{\Delta}{\Omega}=10^{2},\,\alpha=10,$
and $\beta=-2$. Negative regions of the plots in (a)-(d) show antibunching.}
\end{figure}

\end{widetext}

Signatures of nonclassical boson statistics can be obtained in various
ways. Probably, the most popular criterion for detection of nonclassical
boson statistics is $g^{(2)}(0)<1$ where $g^{(2)}(0)$ is the second
order correlation function at zero time delay and it is expressed
as \begin{equation}
\begin{array}{lcl}
g_{j}^{(2)}(0) & = & 1+\frac{D_{j}}{\left\langle N_{j}(t)\right\rangle ^{2}}\end{array}\label{eq:photon statistics}\end{equation}
with \begin{equation}
D_{j}=\left(\Delta N_{j}(t)\right)^{2}-\left\langle N_{j}(t)\right\rangle ,\label{eq:sub-pois}\end{equation}
where $j=\{a,b\}$. The condition $g^{(2)}(0)<1$ is equivalent to
$D_{j}<0$, and it implies sub-Poissonian boson statistics. However,
it is often referred to as condition of antibunching \cite{C T Lee,antibunching-g20}.
We have followed the same convention here, and used $D_{j}<0$ as
the criterion of antibunching. Further, the condition of intermodal
antibunching is \cite{pathak-pra2}\begin{equation}
\begin{array}{lcl}
g_{ab}^{(2)}(0) & = & 1+\frac{D_{ab}}{\left\langle N_{a}(t)\right\rangle \left\langle N_{b}(t)\right\rangle }<1,\end{array}\end{equation}
where $D_{ab}=(\Delta N_{ab})^{2}.$ As $\left\langle N_{j}(t)\right\rangle $
is a nonnegative quantity $g_{ab}^{(2)}(0)<1$ implies $D_{ab}<0$
and vise versa. Consequently, the condition of antibunching as well
as the sub-Poissonian boson statistics for the couple mode is usually
expressed as \begin{equation}
\begin{array}{lcl}
D_{ab} & = & (\Delta N_{ab})^{2}\\
 & = & \left\langle a^{\dagger}(t)b^{\dagger}(t)b(t)a(t)\right\rangle -\left\langle a^{\dagger}(t)a(t)\right\rangle \left\langle b^{\dagger}(t)b(t)\right\rangle <0.\end{array}\label{eq:antib2}\end{equation}
Now using the solution reported here and considering $|\alpha\rangle|\beta\rangle$
as the initial input state, it is easy to obtain closed form analytic
expressions of $D_{j}$ and $D_{ab}$ as follows\begin{equation}
\begin{array}{lcl}
D_{a} & = & \left|f_{2}\right|^{2}\left(\left|\beta\right|^{2}+6\left|\alpha\right|^{2}\left|\beta\right|^{2}-\frac{1}{2}\left|\alpha\right|^{4}\right)\\
 & + & \left\{ \left(f_{2}^{*}f_{1}+f_{5}^{*}f_{1}+f_{2}^{*}f_{3}\right)\alpha^{2}\beta^{*}\right.\\
 & + & \left(f_{6}^{*}f_{1}+f_{2}^{*}f_{4}+4\left|f_{2}\right|^{2}f_{2}^{*}f_{1}\right)\left|\beta\right|^{2}\alpha^{2}\beta^{*}\\
 & + & \left.6\left(f_{5}^{*}f_{1}+f_{2}^{*}f_{3}\right)\left|\alpha\right|^{2}\alpha^{2}\beta^{*}+{\rm c.c.}\right\} ,\end{array}\label{eq:ant-a}\end{equation}
\textcolor{red}{{} }\begin{equation}
\begin{array}{lcl}
D_{b} & = & 2\left(f_{1}^{2}g_{6}\left|\beta\right|^{2}\alpha^{2}\beta^{*}+{\rm c.c.}\right)\end{array}\label{eq:ant-b}\end{equation}
and \begin{equation}
\begin{array}{lcl}
D_{ab} & = & -\left|f_{2}\right|^{2}\left|\alpha\right|^{2}\left|\beta\right|^{2}-\left\{ 2f_{1}^{2}g_{6}\left|\beta\right|^{2}\alpha^{2}\beta^{*}\right.\\
 & + & \left.\left(f_{5}^{*}f_{1}+f_{2}^{*}f_{3}\right)\left|\alpha\right|^{2}\alpha^{2}\beta^{*}+{\rm c.c.}\right\} .\end{array}\label{eq:ant-ab}\end{equation}
Using (\ref{eq:ant-a})-(\ref{eq:ant-ab}), we can easily investigate
the signatures of nonclassical boson statistics in atomic $a$, molecular
$b$, and compound $ab$ modes of the atom-molecule BEC of our interest.
Specifically, we have clearly observed antibunching in molecular mode
$b$ and compound mode $ab,$ for $\alpha=10,$ $\beta=2$ for reasonably
large regions (See Fig. \ref{fig:statistics} b-c), but we have observed
nonclassical boson statistics in atomic mode $a$ only in a small
region for this specific choice of $\alpha,\beta$ (See Fig. \ref{fig:statistics}
a). Further, for $\alpha\leq6$ we have not observed nonclassical
boson statistics in atomic mode. However, antibunching (nonclassical
boson statistics) for all values of rescaled interaction time $\Omega t\,>0$
in atomic mode $a$ for $\alpha\leq6$ can be observed by modifying
the phase of the input coherent state of the molecular mode $b$ (i.e.,
using $\beta=-2$ instead of $\beta=2$). Using $\beta=-2$, we can
also observe antibunching in the molecular mode $b$ for those values
of rescaled interaction time for which we could not observe it for
our previous choice of $\beta$ (i.e., for $\beta=2)$. When the phase
of the molecular mode $b$ is modified (i.e., if we use $\beta=-2$
instead of $\beta=2$) then the phase of all the terms except the
first term in Eq. (\ref{eq:ant-a}) is changed as the first term $\left(\left|f_{2}\right|^{2}\left(\left|\beta\right|^{2}+6\left|\alpha\right|^{2}\left|\beta\right|^{2}-\frac{1}{2}\left|\alpha\right|^{4}\right)\right)$
contains only $\left|\beta\right|$, it is independent of phase of
$\beta$. To be precise, for $\frac{\left|\alpha\right|^{4}}{\left(1+6\left|\alpha\right|^{2}\right)}>2\left|\beta\right|^{2},$
this term gives finite negative contribution to the antibunching in
atomic mode $a$ (as for $\alpha=10,\beta=2$ we obtain $\frac{\left|\alpha\right|^{4}}{\left(1+6\left|\alpha\right|^{2}\right)}=16.6>2\left|\beta\right|^{2}=8)$.
Now if we obtain $D_{a}>0$ for some values of $\alpha$ and $\beta$
(say $\alpha=10,\beta=2$), then that would mean that the sum of all
the terms other than the first term is positive and its absolute value
is greater than that of the first term. Interestingly, this sum can
be made negative by changing the phase of $\beta$ (say by choosing
$\beta=-2$) as it is directly proportional to $\beta$ and clearly
that choice will lead to $D_{a}<0$ or antibunching in mode $a.$
Further, even when $D_{a}<0$ for some values of $\alpha$ and $\beta$
that satisfy $\frac{\left|\alpha\right|^{4}}{\left(1+6\left|\alpha\right|^{2}\right)}>2\left|\beta\right|^{2},$
and the sum of all the terms other than the first term is positive,
but its amplitude is smaller than that of the first term, then by
changing phase of $\beta$ we can increase the depth of nonclassicality.
In brief, by controlling the phase of the input coherent state of
the molecular mode, we can control the nonclassical properties of
the atomic mode $a$. Plots of the analytic results described by Eqs.
(\ref{eq:ant-a}) -(\ref{eq:ant-ab}) are shown in Fig. \ref{fig:statistics}
a-c. Fig. \ref{fig:statistics} d illustrates that the antibunching
in atomic mode $a$ can be observed by modifying the phase of the
input coherent state of the molecular mode.

\subsection{Higher-order antibunching\label{sub:Higher-order-antibunching}}

In order to investigate the higher-order antibunching of the pure
modes, Lee introduced the criterion \cite{C T Lee}\begin{equation}
\begin{array}{lcl}
R(l,m) & = & \frac{\left\langle N_{x}^{\left(l+1\right)}\right\rangle \left\langle N_{x}^{\left(m-1\right)}\right\rangle }{\left\langle N_{x}^{l}\right\rangle \left\langle N_{x}^{m}\right\rangle }\end{array}-1<0,\label{eq:higher order ant1}\end{equation}
where $N$ is the usual number operator and $\left\langle N^{\left(i\right)}\right\rangle =\left\langle N\,(N-1)\,.....(N-i+1)\right\rangle $
is the $i^{th}$ factorial moment of the number operator. $l$ and
$m$ are the integers satisfying the condition $1\leq m\leq l.$ The
subscript $x$ denotes the particular mode. Pathak and Garcia simplified
this criterion for $\left(n-1\right)^{th}$ order antibunching \cite{HOAwithMartin}
as \begin{equation}
\begin{array}{lcl}
\left\langle N_{x}^{\left(n\right)}\right\rangle -\left\langle N_{x}\right\rangle ^{(n)} & < & 0\end{array},\label{eq:higherorder ant2}\end{equation}
where $\left\langle N_{x}^{(n)}\right\rangle =\left\langle a^{\dagger n}a^{n}\right\rangle $
is the measure of probability of observing $n$ bosons of the same
mode at a particular point in space time coordinate. Now, for the
atomic mode $a$ we can obtain $\left(n-1\right)^{th}$ order antibunching
as

\begin{figure}
\begin{centering}
\includegraphics[scale=0.35]{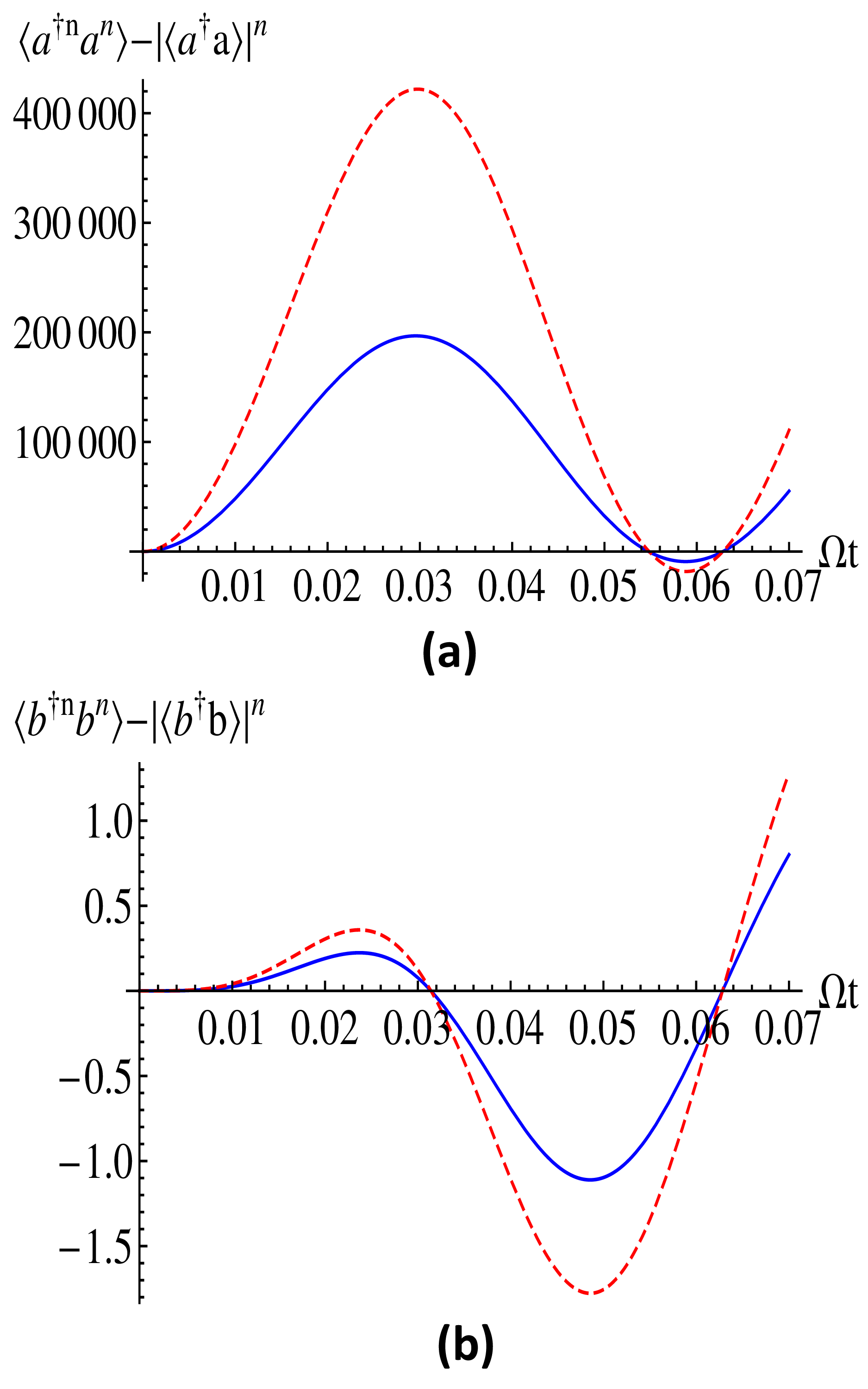}
\par\end{centering}

\caption{\label{fig:Plot-of-HOA}(Color online) Variation of $\left(n-1\right)^{th}$
order antibunching with rescaled time $\Omega t$ for different values
of $n$ for $\Omega=10^{2},\,\frac{\Delta}{\Omega}=10^{2},\,\alpha=10,$
and $\beta=2$ for (a) atomic mode $a$, and (b) molecular mode $b$
with $n=3$ (smooth blue line) and $n=4$ (dashed red line). To display
the plots in the same scale, the Y-axis of the plots for $n=3$ is
amplified by 100 in (a), and by 5 in (b). Negative regions of the
plots show higher-order antibunching in respective modes.}
\end{figure}

\begin{widetext}\begin{equation}
\begin{array}{lcl}
\left\langle a^{\dagger n}a^{n}\right\rangle -\left\langle a^{\dagger}a\right\rangle ^{n} & = & \left|f_{2}\right|^{2}\left\{ \left(^{n}C_{2}\right)^{2}\left|\alpha\right|^{2\left(n-2\right)}\left|\beta\right|^{2}+\left(n^{3}-n\right)\left|\alpha\right|^{2\left(n-1\right)}\left|\beta\right|^{2}-\frac{1}{2}{}^{n}C_{2}\left|\alpha\right|^{2n}\right\} \\
 & + & \left[\left\{ \left(n^{3}-3n^{2}+2n\right)f_{1}^{*2}f_{2}f_{3}+3\left(n^{2}-n\right)f_{1}^{*}f_{5}+\left(n^{3}-n\right)f_{3}^{*}f_{2}\right\} \left|\alpha\right|^{2\left(n-1\right)}\alpha^{*2}\beta\right.\\
 & + & \left\{ ^{n}C_{2}f_{1}^{*}f_{2}+\frac{1}{4}\left(n^{4}-6n^{3}+11n^{2}-6n\right)f_{1}^{*2}f_{2}f_{3}+\left(^{n}C_{2}-6{}^{n}C_{3}\right)f_{1}^{*}f_{5}+\left(^{n}C_{2}\right)^{2}f_{3}^{*}f_{2}\right\} \left|\alpha\right|^{2\left(n-2\right)}\alpha^{*2}\beta\\
 & + & 3^{n}C_{3}f_{1}^{*2}f_{2}^{2}\left|\alpha\right|^{2\left(n-3\right)}\alpha^{*4}\beta^{2}+3^{n}C_{4}f_{1}^{*2}f_{2}^{2}\left|\alpha\right|^{2\left(n-4\right)}\alpha^{*4}\beta^{2}+6^{n}C_{4}f_{1}^{*3}f_{2}^{3}\left|\alpha\right|^{2\left(n-4\right)}\alpha^{*6}\beta^{3}\\
 & + & 15^{n}C_{5}f_{1}^{*3}f_{2}^{3}\left|\alpha\right|^{2\left(n-5\right)}\alpha^{*6}\beta^{3}+15{}^{n}C_{6}f_{1}^{*3}f_{2}^{3}\left|\alpha\right|^{2\left(n-6\right)}\alpha^{*6}\beta^{3}\\
 & + & \left\{ ^{n}C_{2}f_{1}^{*}f_{6}-n\left(n-1\right)^{2}f_{1}^{*2}f_{2}f_{3}+\frac{1}{2}{}^{n}C_{2}\left(3n^{2}-n-4\right)\left|f_{2}\right|^{2}f_{1}^{*}f_{2}-n^{2}\left(n-1\right)f_{3}^{*}f_{2}\right\} \left|\alpha\right|^{2\left(n-2\right)}\left|\beta\right|^{2}\alpha^{*2}\beta\\
 & + & \left.\frac{3}{4}{}^{n}C_{3}n\left(3n-1\right)\left|f_{2}\right|^{2}f_{1}^{*}f_{2}\left|\alpha\right|^{2\left(n-3\right)}\left|\beta\right|^{2}\alpha^{*2}\beta+3{}^{n}C_{2}{}^{n}C_{4}\left|f_{2}\right|^{2}f_{1}^{\star}f_{2}\left|\alpha\right|^{2\left(n-4\right)}\left|\beta\right|^{2}\alpha^{*2}\beta+{\rm c.c.}\right].\end{array}\label{eq:hoa-a}\end{equation}
It is easy to check that for $n=2$, Eq. (\ref{eq:hoa-a}) reduces
to Eq. (\ref{eq:ant-a}). Similarly, for molecular mode, we obtain

\begin{equation}
\left\langle b^{\dagger n}b^{n}\right\rangle -\left\langle b^{\dagger}b\right\rangle ^{n}=\left[n\left(n-1\right)f_{1}^{2}g_{6}\left|\beta\right|^{2\left(n-1\right)}\alpha^{2}\beta^{*}+{\rm c.c.}\right].\label{higherorder-b}\end{equation}
\end{widetext}In order to obtain the flavor of the higher-order nonclassicalities
illustrated by these equations we plot the right hand sides of Eqs.
(\ref{eq:hoa-a}) and (\ref{higherorder-b}) in Fig. \ref{fig:Plot-of-HOA}
a-b. Clearly the figures illustrate the existence of higher-order
antibunching, and also show that the depth of nonclassicality increases
with the increase in the order of antibunching i.e., $n$. This is
consistent with the earlier observations \cite{HOAwithMartin,HOAis not rare}
and it shows that the detection of weaker nonclassicality becomes
easier when a higher-order criterion is used, as shown in \cite{Maria-PRA-1,Maria-2}.
It is also worth mentioning here that higher-order antibunching in
the atomic mode $a$ and molecular mode $b$ can be controlled by
controlling the phase of the input coherent state in the molecular
mode $b$. For example, if we use $\beta=-2$ instead of $\beta=2$
in (\ref{higherorder-b}) then we observe higher-order antibunching
for values of rescaled time for which we could not observe higher-order
antibunching with the previous choice (i.e., for $\beta=2$).\textcolor{green}{{}
}%

\section{{\normalsize Entanglement\label{sec:Entanglement}}}

In order to study the two-mode entanglement, we use the Hillery-Zubairy
(HZ) criteria (1 and 2) \cite{HZ-PRL,HZ2007,HZ2010} and Duan et al.'s
criterion\textcolor{blue}{{} }\cite{duan}. Lower-order HZ-1 and HZ-2
criteria are expressed as \begin{equation}
\left\langle N_{a}(t)N_{b}(t)\right\rangle -\left|\left\langle a(t)b^{\dagger}(t)\right\rangle \right|^{2}<0\label{eq:hz1-crit}\end{equation}
and \begin{equation}
\left\langle N_{a}(t)\right\rangle \left\langle N_{b}(t)\right\rangle -\left|\left\langle a(t)b(t)\right\rangle \right|^{2}<0,\label{eq:hz2-crit}\end{equation}
respectively. Using (\ref{eq:evolution}), (\ref{eq:solution}), (\ref{eq:coherent state})
and (\ref{eq:hz1-crit}), we obtain

\begin{figure}
\begin{centering}
\includegraphics[scale=0.35]{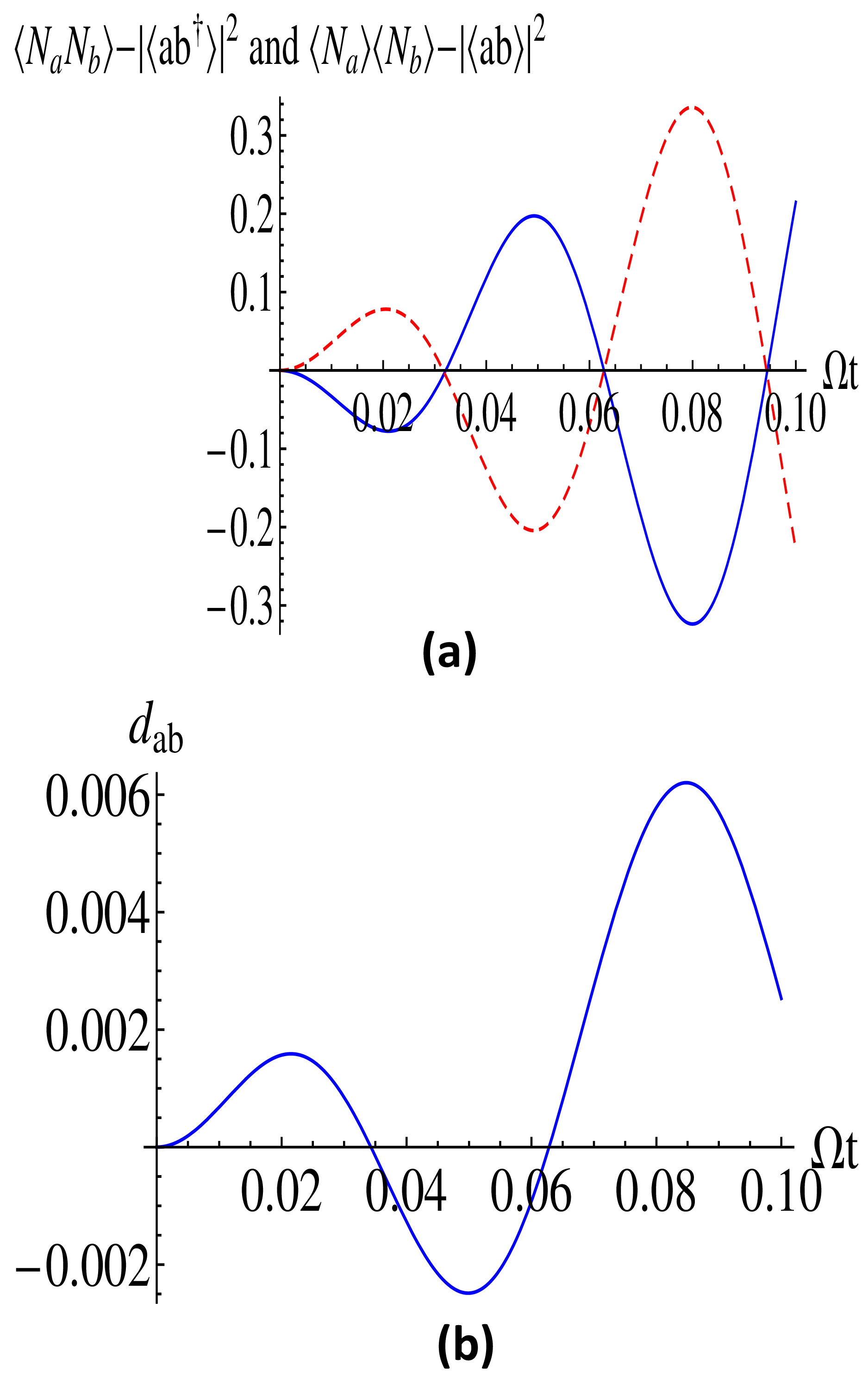}
\par\end{centering}

\caption{\label{fig:ent}(Color online) Plots show variation of intermodal
entanglement in mode $ab$ with rescaled interaction time $\Omega t$
using (a) Hillery-Zubairy criterion-1 (HZ-1) smooth blue line and
Hillery-Zubairy criterion-2 (HZ-2) dashed red line and (b) Duan criterion
for $\Omega=10^{2},\,\frac{\Delta}{\Omega}=10^{2},\,\alpha=10,$ and
$\beta=2$. Negative regions of the plots in (a) and (b) show intermodal
entanglement. (a) shows atomic and molecular modes are always entangled
for this particular choice of $\alpha$ and $\beta$.}
\end{figure}

\begin{widetext}\begin{equation}
\begin{array}{lcl}
\left\langle N_{a}(t)N_{b}(t)\right\rangle -\left|\left\langle a(t)b^{\dagger}(t)\right\rangle \right|^{2} & = & \left|f_{2}\right|^{2}\left(\left|\beta\right|^{4}-\left|\alpha\right|^{2}\left|\beta\right|^{2}\right)+\left\{ \left(3f_{2}^{*}f_{3}^{*}f_{1}^{2}+4f_{2}^{*}f_{3}-g_{7}\right)\left|\beta\right|^{2}\alpha^{2}\beta^{*}\right.\\
 & - & \left.\left(f_{5}^{*}f_{1}+f_{2}^{*}f_{3}\right)\left|\alpha\right|^{2}\alpha^{2}\beta^{*}+{\rm c.c.}\right\} ,\end{array}\label{eq:hz1}\end{equation}
Similarly, using (\ref{eq:evolution}), (\ref{eq:solution}), (\ref{eq:coherent state})
and (\ref{eq:hz2-crit}), we obtain \begin{equation}
\begin{array}{lcl}
\left\langle N_{a}(t)\right\rangle \left\langle N_{b}(t)\right\rangle -\left|\left\langle a(t)b(t)\right\rangle \right|^{2} & = & \left|f_{2}\right|^{2}\left(\left|\beta\right|^{4}+\left|\alpha\right|^{2}\left|\beta\right|^{2}\right)+\left\{ \left(2f_{5}^{*}f_{1}-f_{2}^{*}f_{3}^{*}f_{1}^{2}\right)\left|\beta\right|^{2}\alpha^{2}\beta^{*}\right.\\
 & - & \left.\left(g_{8}+g_{4}^{*}g_{2}\right)\left|\alpha\right|^{2}\alpha^{2}\beta^{*}+{\rm c.c.}\right\} .\end{array}\label{eq:hz2}\end{equation}
\end{widetext}As both HZ-1 and HZ-2 criteria are only sufficient
and not essential, we also investigate the existence of entanglement
using Duan et al.'s criterion \cite{duan} which can be written as
\begin{equation}
\begin{array}{lcl}
d_{ab}=\left\langle \left(\triangle u_{ab}\right)^{2}\right\rangle +\left\langle \left(\triangle v_{ab}\right)^{2}\right\rangle -2 & < & 0\end{array},\label{eq:duan crit}\end{equation}
where \begin{equation}
\begin{array}{lcl}
u_{ab} & = & \frac{1}{\sqrt{2}}\left\{ \left(a+a^{\dagger}\right)+\left(b+b^{\dagger}\right)\right\} ,\\
v_{ab} & = & -\frac{i}{\sqrt{2}}\left\{ \left(a-a^{\dagger}\right)+\left(b-b^{\dagger}\right)\right\} .\end{array}\label{eq:duan-2}\end{equation}
Here we would like to note that all the inseparability criteria described
above and in the rest of the paper can be obtained as special cases
of Shchukin-Vogel entanglement criterion \cite{vogel's ent criterion}.
Using Eqs. (\ref{eq:evolution}), (\ref{eq:solution}), (\ref{eq:coherent state})
and (\ref{eq:duan crit}), we obtain \begin{equation}
\begin{array}{lcl}
d_{ab} & = & 2\left\{ \left|f_{2}\right|^{2}\left|\beta\right|^{2}+\left(f_{2}^{*}f_{3}\alpha^{2}\beta^{*}+f_{2}g_{4}^{*}|\beta|^{2}\alpha^{*}+{\rm c.c.}\right)\right\} .\end{array}\label{eq:duan}\end{equation}
Temporal evolution of the parameters that indicate the existence of
entanglement are shown in the Fig. \ref{fig:ent} a using HZ-1 and
HZ-2 criterion and in the Fig. \ref{fig:ent} b using Duan et al.'s
criteria, respectively. Fig. \ref{fig:ent} a shows that the atomic
and molecular modes are entangled for any value of $\Omega t\,>0$
for the specific values of parameters chosen here. 

\subsection{Higher-order entanglement}

In order to investigate the higher-order entanglement for the coupled
mode $ab,$ we use the two criteria of Hillery-Zubairy \cite{HZ-PRL}.
These are\begin{equation}
\begin{array}{lcl}
\left\langle a^{\dagger n}a^{n}b^{\dagger m}b^{m}\right\rangle -\left|\left\langle a^{n}b^{\dagger m}\right\rangle \right|^{2} & < & 0\end{array}\label{higerorderHZ1}\end{equation}
and \begin{equation}
\left\langle a^{\dagger n}a^{n}\right\rangle \left\langle b^{\dagger m}b^{m}\right\rangle -\left|\left\langle a^{n}b^{m}\right\rangle \right|^{2}<0,\label{higherorderHZ2}\end{equation}
where $m$ and $n$ are positive integers and for higher-order entanglement
$m+n\geq3$. It is easy to observe that for $m=1$ and $n=1,$ criteria
(\ref{higerorderHZ1}) and (\ref{higherorderHZ2}) reduces to (\ref{eq:hz1-crit})
and (\ref{eq:hz2-crit}), respectively. This is why criteria (\ref{higerorderHZ1})
and (\ref{higherorderHZ2}) are usually referred to as higher-order
HZ-1 and HZ-2 criteria, respectively. Now using Eqs. (\ref{eq:solution}),
(\ref{eq:coherent state}) and criterion (\ref{higerorderHZ1}) for
a specific case $n=1,\, m=2$, we obtain 

\begin{figure}
\begin{centering}
\includegraphics[scale=0.35]{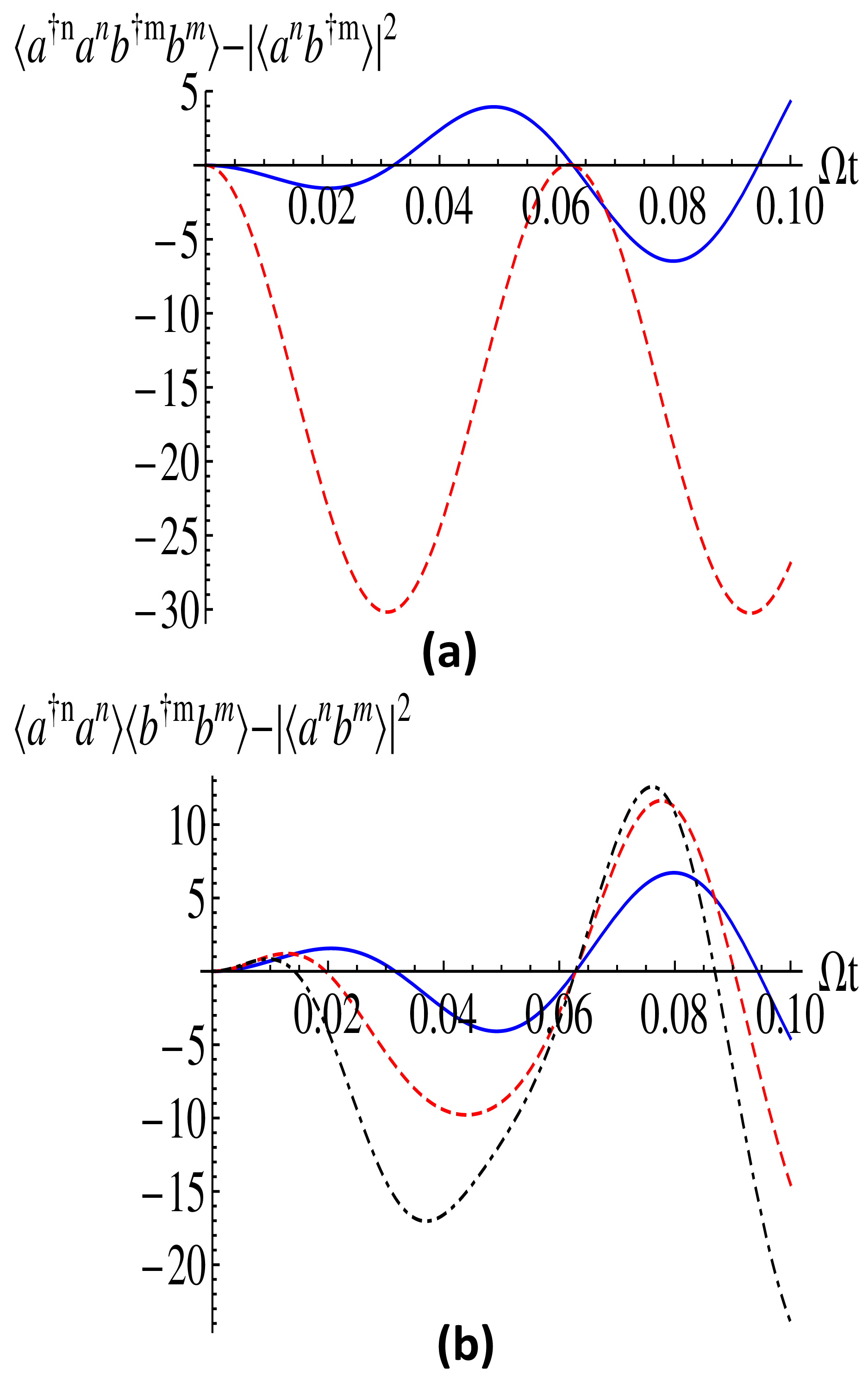}
\par\end{centering}

\caption{\label{fig:Plot-of-Higher-ent}(Color online) Variation of higher-order
intermodal entanglement parameters ($\left\langle a^{\dagger n}a^{n}b^{\dagger m}b^{m}\right\rangle -\left|\left\langle a^{n}b^{\dagger m}\right\rangle \right|^{2}$
and $\left\langle a^{\dagger n}a^{n}\right\rangle \left\langle b^{\dagger m}b^{m}\right\rangle -\left|\left\langle a^{n}b^{m}\right\rangle \right|^{2}$)
with rescaled time $\Omega t$ for $\Omega=10^{2},\,\frac{\Delta}{\Omega}=10^{2},\,\alpha=10,$
and $\beta=2$ using (a) higher-order HZ-1 criterion for $m=n=1$
(smooth blue line) and $m=2$ and $n=1$ (dashed red line) (b) HZ-2
criterion for $m=1$ and $n=1$ (smooth blue line), $m=2$,$n=1$
(dashed red line) and $m=3,n=1$ (dot-dashed black line). To show
the plots in the same scale in (a) and (b) the $Y$-axis of plots
for lower-order entanglement (i.e., the smooth line in (a) and (b))
is amplified by a factor of 20, and for $m=2$ in (b) by a factor
of 5. Negative regions of plots for $m+n\geq3$ (i.e., dotted and
dot-dashed lines in (a) and (b)) show the existence of higher-order
entanglement.}
\end{figure}

\begin{widetext}

\begin{equation}
\begin{array}{lcl}
\left\langle a^{\dagger}ab^{\dagger2}b^{2}\right\rangle -\left|\left\langle ab^{\dagger2}\right\rangle \right|^{2} & = & \left|f_{2}\right|^{2}\left(\left|\beta\right|^{6}-2\left|\alpha\right|^{4}\left|\beta\right|^{2}\right)+\left\{ \left(f_{2}^{*}f_{3}+6\left|f_{2}\right|^{2}g_{2}-2g_{7}\right)\left|\beta\right|^{4}\alpha^{2}\beta^{*}\right.\\
 & - & \left.\left(2f_{2}^{*}f_{3}+2\left|f_{2}\right|^{2}g_{2}+g_{7}\right)\left|\alpha\right|^{2}\left|\beta\right|^{2}\alpha^{2}\beta^{*}+{\rm c.c.}\right\} .\end{array}\label{eq:higher hz1}\end{equation}
Right hand side of (\ref{eq:higher hz1}) is plotted in Fig. \ref{fig:Plot-of-Higher-ent}
a and we can clearly see the existence of higher-order entanglement
through the negative regions of the plot. In this case we have investigated
the existence of higher-order entanglement using a particular value
of $m$ and $n,$ but it is possible to obtain a general expression
for arbitrary values of $m$ and $n$ using the present framework.
Just to illustrate this point, we use Eqs. (\ref{eq:evolution}),
(\ref{eq:solution}), (\ref{eq:coherent state}) and criterion (\ref{higherorderHZ2})
in general to obtain\begin{equation}
\begin{array}{lcl}
\left\langle a^{\dagger n}a^{n}\right\rangle \left\langle b^{\dagger m}b^{m}\right\rangle -\left|\left\langle a^{n}b^{m}\right\rangle \right|^{2} & = & \left|f_{2}\right|^{2}\left(mn\left|\alpha\right|^{2n}\left|\beta\right|^{2m}+n^{2}\left|\alpha\right|^{2\left(n-1\right)}\left|\beta\right|^{2\left(m+1\right)}\right)\\
 & + & \left\{ mn\left(m-1\right)f_{1}^{*2}f_{2}f_{3}+mnf_{1}^{*}f_{5}+m^{2}nf_{3}^{*}f_{2}\right\} \left|\alpha\right|^{2n}\left|\beta\right|^{2\left(m-1\right)}\alpha^{*2}\beta\\
 & + & \left\{ 2mnf_{1}^{*}f_{5}+n^{2}\left(1-2m\right)f_{3}^{*}f_{2}-\frac{1}{2}mn^{2}\left|f_{2}\right|^{2}f_{1}^{*}f_{2}\right.\\
 & - & \left.2mn^{2}f_{1}^{*2}f_{2}f_{3}\right\} \left|\alpha\right|^{2\left(n-1\right)}\left|\beta\right|^{2m}\alpha^{*2}\beta\\
 & - & n\left(n-1\right)\left\{ mf_{8}^{*}f_{1}+\left(n-2\right)mf_{3}^{*}f_{2}+mnf_{1}^{*2}f_{2}f_{3}\right\} \left|\alpha\right|^{2\left(n-2\right)}\left|\beta\right|^{2m}\alpha^{*2}\beta\\
 & + & n^{2}\left(n-1\right)\left|f_{2}\right|^{2}f_{1}^{*}f_{2}\left\{ \left|\alpha\right|^{2\left(n-2\right)}\left|\beta\right|^{2\left(m+1\right)}\alpha^{*2}\beta\right.\\
 & + & \left.\left.\frac{1}{2}\left(n-2\right)\left|\alpha\right|^{2\left(n-3\right)}\left|\beta\right|^{2\left(m+1\right)}\alpha^{*2}\beta\right\} +{\rm c.c.}\right].\end{array}\label{eq:higherhz2}\end{equation}

\end{widetext}

In Fig. \ref{fig:Plot-of-Higher-ent} b we show the existence of the
higher-order entanglement in the atom-molecule BEC using the higher-order
HZ-2 criterion for various values of $m$ with $n=1$. As before,
we observed that the depth of nonclassicality increases with the increase
in order of entanglement.

\textcolor{red}{}%

\section{{\normalsize Conclusions\label{sec:Conclusions}}}

Lower-order and higher-order nonclassical properties of a two-mode
atom-molecule BEC system is investigated here with the help of a third
order perturbative solution of the Heisenberg's equations of motion
corresponding to the Hamiltonian of the BEC system. The investigation
established that even if we start with classical (i.e., coherent and
separable) input state, the interaction introduces nonclassicality.
Thus, the interaction between the atomic and molecular modes leads
to a superposition in tensor product space and phase space. Specifically,
we considered a separable initial state as we assumed it to be a product
of two coherent states ($|\alpha\rangle|\beta\rangle$). Now, using
lower-order, and higher-order inseparability criteria of Hillery and
Zubairy, and Duan et al.'s criterion, we have shown the existence
of lower-order, and higher-order intermodal entanglement. As an entangled
state can always be viewed as a superposition of separable states
in the tensor product space (for example, Bell state is an equal superposition
of $|0\rangle\otimes|0\rangle$ and $|1\rangle\otimes|1\rangle$),
we may conclude that the interaction between the atomic mode and molecular
mode of the two-mode BEC system leads to a superposition in tensor
product space and the existence of this superposition in tensor product
space is reflected here when we observed entanglement through the
inseparability criteria mentioned above. In a similar fashion, a traditional
nonclassical state such as squeezed state or a Fock state may be viewed
as superposition of coherent states \cite{vogel-ent meas}. Thus,
the lower-order, and higher-order squeezing and antibunching observed
here is essentially a manifestation of superposition in phase (Hilbert)
space due to interaction between atomic and molecular modes. Interestingly,
amount of superposition (i.e., interference in phase space and/or
tensor product space) can be controlled by controlling various parameters,
such as, interaction time, coupling constant $\Omega$, and detuning
$\Delta$, boson number of the input modes and the phase of the input
coherent states. The effects of these parameters are illustrated through
Figs. \ref{fig:squeezing}-\ref{fig:Plot-of-Higher-ent}. 

The methodology adopted here and in our earlier work on atom-atom
two-mode BEC \cite{pathak-pra2}\textcolor{red}{{} }is quite general,
and is applicable to other bosonic systems, too. Further, in Table
I and II of Ref. \cite{Adam-criterion}, a large number of nonclassical
criteria based on the expectation values of the moments of the annihilation
and creation operators of the field modes is listed. As we already
have compact expressions for the field operators, it is possible to
extend the present work to investigate other signatures of nonclassicality
using the criteria of nonclassicality listed in \cite{Adam-criterion}.
For example, we can easily extend the present work to study hyperbunching
\cite{hyperbunching}, sum and difference squeezing of An-Tinh \cite{sum&difference-sq-An-Tinh}
and Hillery \cite{Sum&diff-sq-Hillery}, the existence of entanglement
using the inseparability criterion of Manicini et al. \cite{Mancini-ent},
Simon \cite{Simon} and Miranowicz et al. \cite{Adam-ent}, etc. In
addition, recent experimental successes in realizing two-mode BEC
systems and observing higher-order nonclassicality indicate that the
observations of the present theoretical work can be verified experimentally.
We conclude the paper with a hope that the results presented in this
work will be useful in the future development of quantum information
processing in particular and nonclassical states in general.

\textbf{Acknowledgment:} A. P. and K. T. thank Department of Science
and Technology (DST), India for support provided through the DST project
No. SR/S2/LOP-0012/2010.

\end{document}